\documentclass[epj]{svjour}
\usepackage{latexsym}
\usepackage{epsfig}
\usepackage{float}
\usepackage{lipsum}
\usepackage{graphicx,times}
\usepackage{amssymb,amsmath}
\usepackage{color}
\usepackage{dcolumn}
\usepackage{epstopdf}
\usepackage{pgfplots}
\usepackage[colorlinks=true, urlcolor=blue, linkcolor=red, citecolor=blue]{hyperref}
\begin{document}
\title{Trace of $\Lambda(1405)$ resonance in low energy 
$K^{-}+\, ^{3}\mathrm{He}\rightarrow (\pi^{0}\Sigma^{0})+d$ reaction}
\author{J. Esmaili\inst{1}, S. Marri\inst{2}, M. Raeisi\inst{1} and A. Naderi Beni\inst{3}}
\institute{Department of Physics, Faculty of Basic Sciences, Shahrekord University, 
Shahrekord, 115, Iran\and Department of Physics, Isfahan University of Technology, 
Isfahan 84156-83111, Iran \and Department of Physics, Payame Noor University, P. O. 
Box 19395-3697, Tehran, Iran}
\date{Received: date / Revised version: date}
\abstract{
In present work, we investigated $K^{-}+\, ^{3}\mathrm{He}$ reaction at low energies. The 
coupled-channel Faddeev AGS equations were solved for $\bar{K}Nd-\pi\Sigma{d}$ three-body 
system in momentum representation to extract the scattering amplitudes. To trace the signature 
of the $\Lambda$(1405) resonance in the $\pi\Sigma$ invariant mass, the deuteron energy 
spectrum for $K^{-}+\,^{3}\mathrm{He}\rightarrow\pi\Sigma{d}$ reaction was obtained. Different 
types of $\bar{K}N-\pi\Sigma$ potentials based on phenomenological and chiral SU(3) approaches 
were used. As a remarkable result of this investigation, it was found that the deuteron energy 
spectrum, reflecting the $\Lambda$(1405) mass distribution and width, depends quite sensitively 
on the $\bar{K}N-\pi\Sigma$ model of interaction. Hence accurate measurements of the $\pi\Sigma$ 
mass distribution have the potential to discriminate between possible mechanisms at 
work in the formation of the $\Lambda$(1405).
\PACS{{13.75.Jz, 14.20.Pt, 21.85.+d, 25.80.Nv}{describing text of that key}}
} 
\maketitle
\section{Introduction}
\label{intro}
An important issue in various aspects in strangeness nuclear physics is the structure of 
$\Lambda(1405)$ resonance, which has been found to be a highly controversies topic in 
studying the antikaon-nucleon interaction. The $\Lambda(1405)$ resonance is a bound state 
of $\bar{K}N$, which exclusively decays into the $\pi\Sigma(I=0)$ channel via the strong 
interaction. The $\bar{K}N-\pi\Sigma$ interaction, which is a fundamental ingredient of 
the antikaonic nuclear clusters~\cite{ak1,ak2,ak3,ak4,sh1,sh2,ik1,ik2,do1,do2,ik3,kh1,kh2,kh3,kh4,kh5} 
is also strongly dominated by $\Lambda(1405)$ resonance. The existence of $\Lambda(1405)$ 
resonance was first predicted by Dalitz and Tuan~\cite{25,26} in 1959 showing that the 
unitarity in coupled-channel $\bar{K}N-\pi\Sigma$ system leads to the existence of 
$\Lambda(1405)$. As early as in 1961 an experimental evidence of this resonance was 
reported in the invariant mass spectrum of the $\pi\Sigma$ resulting from 
$K^{-}p\rightarrow\pi\pi\pi\Sigma$ reaction~\cite{27} at 1.15 GeV.

The $\bar{K}N$ interaction models which reproduce the mass of $\Lambda(1405)$ resonance 
and two-body scattering data can be divided into two classes: those constructed 
phenomenologically~\cite{sh1,sh2,208} and those derived based on chiral SU(3) dynamics
~\cite{ik1,ik2,do1,do2,ik3}. Even though the phenomenological and the chiral SU(3) 
$\bar{K}N$ interaction models produce comparable results at and above $\bar{K}N$ 
threshold, they differ significantly in their extrapolations to sub-threshold energies
~\cite{hy1}. The phenomenological $\bar{K}N$ potentials are constructed to describe 
the $\Lambda(1405)$ as a single pole of the scattering amplitude around 1405 MeV, 
corresponding to a quasi-bound state of the $\bar{K}N$ system with a binding energy 
of about 30 MeV. On the other hand, the $\bar{K}N-\pi\Sigma$ coupled-channels amplitude 
resulting from chiral SU(3) dynamics has two poles. The two poles are commonly characterized 
as following: the first pole in the complex energy plane is located quite close to the 
$\bar{K}N$ threshold with a small imaginary part, around 10-30 MeV, and a strong coupling 
to $\bar{K}N$. In turn, the second one is wider, with a relatively large imaginary part 
around 50-200 MeV, coupling more strongly to the $\pi\Sigma$ channel and its pole position 
shows more dependence on the specific theoretical model~\cite{os1,a11,a13,a14,ol1,ra1}. 
This different pole structure comes from different off-shell properties of the $\bar{K}N$ 
interactions. The $\bar{K}N$ interactions based on the chiral SU(3) dynamics are 
energy-dependent, and that in the sub-threshold become less attractive than those 
proposed by the energy-independent phenomenological potentials~\cite{hy1}.

The $\pi\Sigma$ mass spectrum is a suitable tools to study the $\bar{K}N$ reaction below 
the $\bar{K}N$ threshold. As it is impossible to perform the scattering experiment in the 
$\pi\Sigma$ channel directly, the resonance properties can be extracted by analyzing the 
invariant mass distribution of the $\pi\Sigma$ final state in reactions that produce 
$\Lambda(1405)$ resonance. During the past decades, many experimental and theoretical 
searches were carried out to investigate the possible observation of $\Lambda(1405)$ 
resonance. Braun {\it et al.,} studied the $K^{-}d$ reaction and reported a resonance 
energy around 1420 MeV~\cite{28}. In Ref.~\cite{29}, the pion induced reaction 
$(\pi^{-}p\rightarrow{K^{+}}\pi\Sigma)$ was investigated and the mass of the resonance 
was found to be consistent with 1405 MeV. Using photo-production reactions, the CLAS
~\cite{mor1,mor2,mor3} and LEPS~\cite{37,38} collaboration investigated the $\Lambda(1405)$ 
resonance signal. Several theoretical studies have been done to analyze the CLAS data 
using different interaction models for $\bar{K}N$ system, which are based on chiral 
SU(3) dynamics~\cite{roca1,roca2,naka,mai,mis} and phenomenological approachs~\cite{has}. 
Other interesting experiments were also performed at GSI by HADES collaboration~\cite{41} 
and at J-PARC as an E31 experiment~\cite{47} to clarify the nature of the $\Lambda(1405)$ 
by using $pp$ and $K^{-}d$ reactions, respectively. In the E31 experiment, the $\pi\Sigma$ 
mass spectra are measured for all combinations of charges, i.e., $\pi^{\pm}\Sigma^{\mp}$ 
and $\pi^{0}\Sigma^{0}$. To establish a theoretical framework for a detailed analysis of 
$K^{-}d$ reaction, different theoretical studies were performed. The theoretical investigations 
of $K^{-}d\rightarrow\pi\Sigma+n$ have been performed in~\cite{ec1,ec2,ec3,ec4} using a 
two-step process. The differential cross sections of $K^{-}d$ reaction were also studied 
in Refs.~\cite{ec5,ec6} using three-body Faddeev method and it was demonstrated that the 
$K^{-}d$ reaction can be a useful tool for studying the sub-threshold properties of the 
$\bar{K}N$ interaction.

The $K^{-}+\,^{3}\mathrm{He}$ reaction was studied in Ref.~\cite{esm1} using variational 
method. The $\Sigma\pi$ invariant-mass spectrum in the resonant capture of $K^{-}$ at rest 
in $^3\mathrm{He}$ were calculated by a coupled-channel potential for $\bar{K}N-\pi\Sigma$ 
interaction. The results in Ref.~\cite{esm1} confirmed the $\Lambda(1405)$ ansatz and the 
proposed predictions by chiral-SU(3) ($M \sim 1420$ MeV/$c^2$) were excluded. The authors 
finally proposed more stringent test by using a $^3$He target. An experimental search for 
$\bar{K}NN$ bound state was performed at J-PARC by using the in-flight $K^{-}+\,^{3}\mathrm{He}$ 
reaction at 1 GeV/c~\cite{sak}. In the E15 experiment, the $\pi\Sigma$ invariant mass spectrum 
resulting from $K^{-}+\,^{3}\mathrm{He}$ reaction was measured for two combinations of charges, 
i.e., $\pi^{\pm}\Sigma^{\mp}{pn}$. It was shown that the production cross section of the 
$\Lambda(1405)p$ is $\sim 10$ times larger than that of the $K^{-}pp$ bound state observed 
in $\Lambda p$ invariant mass which is an important information on the production mechanism 
of the $K^{-}pp$ bound state~\cite{sak}.

\begin{figure*}[t]
\centering
\includegraphics[width=8.3cm]{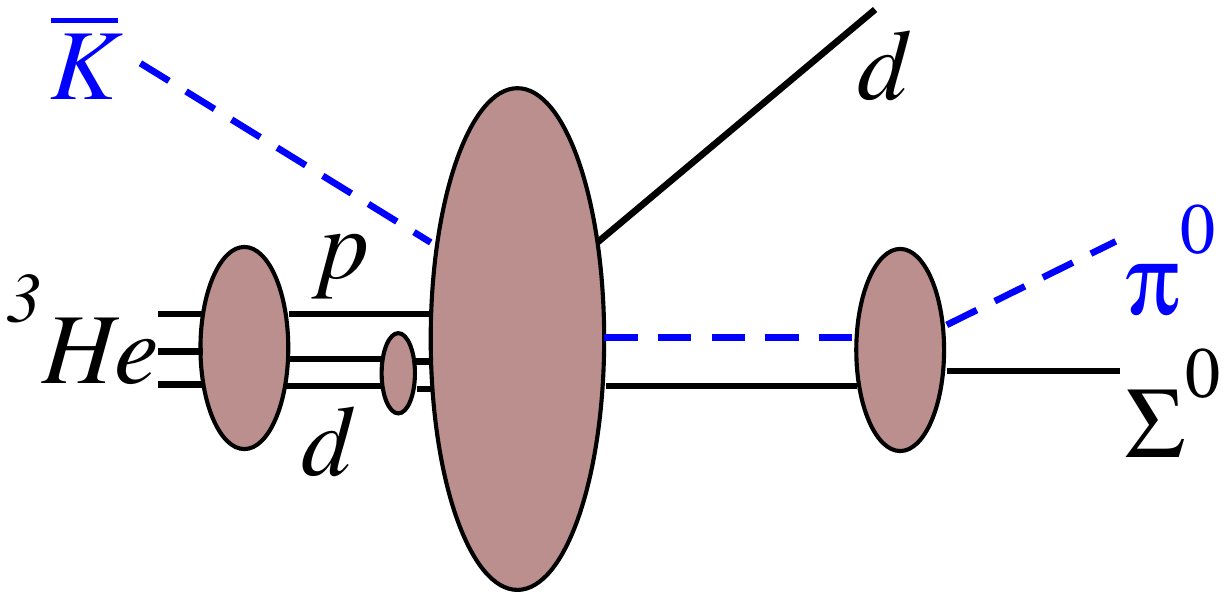}
\caption{(Color on line) Diagrammatic representation for 
$K^{-}+\,^{3}\mathrm{He}\rightarrow(\pi\Sigma)+{d}$ reaction 
using a $p-d$ cluster structure for $^{3}\mathrm{He}$ nuclei.} 
\label{fig.1}
\end{figure*}

The purpose of the present work is to explore the $\pi\Sigma$ invariant mass spectra resulting 
from the $K^{-}+\,^{3}\mathrm{He}\rightarrow\pi\Sigma{d}$ reaction. The problem can be solved 
using methods developed within three-body theories. To reduce the four-body $K^{-}+\,^{3}\mathrm{He}$ 
system to a three-body system, we considered a $p-d$ cluster structure for $^{3}\mathrm{He}$ 
nuclei (Fig.~\ref{fig.1}). To study this reaction, the Faddeev amplitudes for $\bar{K}Nd-\pi\Sigma{d}$ 
system were calculated at real scattering energies. One of the aims is to study the role of 
different off-shell properties of the underlying interactions as they are realized in chiral 
SU(3) dynamics versus phenomenological potential models. With this method, we investigated how 
well the $\Lambda(1405)$ resonance manifests itself in the three-body observable. To study 
the dependence of the $K^{-}+\,^{3}\mathrm{He}$ reaction on the fundamental $\bar{K}N-\pi\Sigma$ 
interaction, different interaction models derived from chiral SU(3) and phenomenological 
approaches, were included in our calculations.

The paper is organized as follows: in Sect.~\ref{formula}, we will explain the Faddeev 
formalism used for the three-body $\bar{K}Nd$ system and give a brief description of 
scattering amplitude for $K^{-}+\,^{3}\mathrm{He}\rightarrow\pi\Sigma{d}$ reaction. The 
two-body inputs of the calculations and the extracted results for $\pi\Sigma$ mass 
spectra are presented in Sect.~\ref{result} and in Section~\ref{conc}, we give conclusions.
\section{Three-body treatment of $K^{-}+\,^{3}\mathrm{He}$ reaction}
\label{formula}
In the present work, the possible signature of the $\Lambda(1405)$ resonance in 
$\pi\Sigma$ mass spectrum resulting from $K^{-}+\,^{3}\mathrm{He}\rightarrow\pi\Sigma{d}$ 
reaction was studied. We used the three-body Faddeev AGS equations~\cite{ags}. As 
there are three different particles in the system under consideration, we will have 
the following partitions of the $\bar{K}Nd$ three-body system, defining the interacting 
pairs and their allowed spin and isospin quantum numbers
\begin{equation}
\begin{split}
& (1):\bar{K}+(Nd)_{s=\frac{1}{2};I=\frac{1}{2}}, \\ 
& (2):N+(\bar{K}d)_{s=1;I=\frac{1}{2}},\\
& (3):d+(\bar{K}N)_{s=\frac{1}{2};I=0}.
\end{split}
\label{eq.1}
\end{equation}

The quantum numbers of the $\bar{K}Nd$ are $I=0$ and $s=\frac{1}{2}$, in actual 
calculations, when we include isospin and spin indices the number of configurations 
is equal to three, corresponding to different possible two-quasi-particle partitions.

The key point of the present calculations is the separable representation of the 
scattering amplitudes in the two-body subsystems. The separable potentials for 
two-body subsystems are given by 
\begin{equation}
V^{I_{i}}_{i}(k,k')=g^{I_{i}}_{i}(k)\, \lambda^{I_{i}}_{i}\, g^{I_{i}}_{i}(k'),
\label{eq.2}
\end{equation}
where $g^{I_{i}}_{i}(k)$ is used to define the form factor of the interacting 
two-body subsystem with relative momentum $k$ and isospin $I$ and $\lambda^{I_{i}}_{i}$ 
defines the strength of the interaction. The two-body interactions are also labeled by 
the $i$ values to define simultaneously the spectator particle and interacting pair. 
Using separable potentials, we can define the two-body t-matrices in the following form
\begin{equation}
T^{I_{i}}_{i}(k,k';z)=g^{I_{i}}_{i}(k)\, 
\tau^{I_{i}}_{i}(z-\frac{p^{2}_{i}}{2\eta_{i}})\, g^{I_{i}}_{i}(k'),
\label{eq.3}
\end{equation}
where the $\tau^{I_{i}}_{i}(z)$-functions are the two-body propagators embedded in 
three-body system and $p_{i}$ is the spectator particle momentum. The reduced mass 
$\eta_{i}$ is also given by 
\begin{equation}
\eta_{i} = m_{i}(m_{j}+m_{k})/(m_{i}+m_{j}+m_{k}).
\label{eq.4}
\end{equation}

\begin{figure*}[t]
\centering
\includegraphics[width=14.cm]{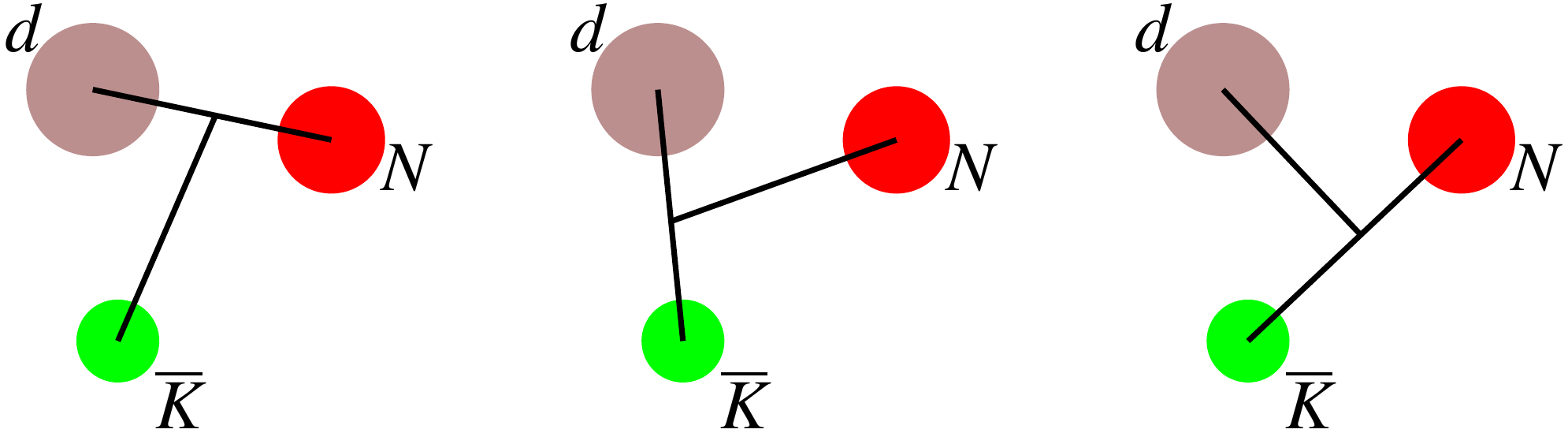}
\caption{(Color on line) Diagrammatic representation for different 
partitions of the $\bar{K}Nd$ system. Defining the interacting 
particles, we will have three partitions, namely $\bar{K}+(Nd)$, 
$(\bar{K}d)+N$ and $(\bar{K}N)+d$. The anti-kaon is defined by 
green circle, the nucleon by red circle and the deuteron by brown circle.} 
\label{fig.2}
\end{figure*}

The whole dynamics of $\bar{K}Nd$ three-body system is described in terms of the 
transition amplitudes $\mathcal{K}_{i,m;j,n}^{I_{i}I_{j}}$, which connect the 
quasi-two-body channels characterized by Eq.~\ref{eq.1}. In Fig.~\ref{fig.2}, the 
three different rearrangement channels of the $\bar{K}Nd$ are represented. The 
Faddeev AGS equations for $\bar{K}Nd$ systems can be expressed by
\begin{equation}
\begin{split}
& \mathcal{K}_{\bar{K},m;\bar{K},n}^{I_{\bar{K}}I_{\bar{K}}}=
 \sum_{rr'}\mathcal{M}_{\bar{K},m;N,r}^{I_{\bar{K}}I_{N}}\tau_{N(rr')}^{I_{N}}
 \mathcal{K}_{N,r';\bar{K},n}^{I_{N}I_{\bar{K}}} \\
& \hspace{1.5cm}+\sum_{rr'}\mathcal{M}_{\bar{K},m;d,r}^{I_{\bar{K}}I_{d}}
\tau_{d(rr')}^{I_{d}}\mathcal{K}_{d,r';\bar{K},n}^{I_{d}I_{\bar{K}}} \\
& \mathcal{K}_{N,m;\bar{K},n}^{I_{N}I_{\bar{K}}}= 
\mathcal{M}_{N,m;\bar{K},n}^{I_{N}I_{\bar{K}}}+ 
 \sum_{rr'}\mathcal{M}_{N,m;\bar{K},r}^{I_{N}I_{\bar{K}}}
\tau_{\bar{K}(rr')}^{I_{\bar{K}}}
\mathcal{K}_{\bar{K},r';\bar{K},n}^{I_{\bar{K}}I_{\bar{K}}} \\
& \hspace{1.5cm}+\sum_{rr'}\mathcal{M}_{N,m;d,r}^{I_{N}I_{d}}\tau_{d(rr')}^{I_{d}}
\mathcal{K}_{d,r';\bar{K},n}^{I_{d}I_{\bar{K}}} \\
& \mathcal{K}_{d,m;\bar{K},n}^{I_{d}I_{\bar{K}}} = 
\mathcal{M}_{d,m;\bar{K},n}^{I_{d}I_{\bar{K}}}+
 \sum_{rr'}\mathcal{M}_{d,m;\bar{K},r}^{I_{d}I_{\bar{K}}}
\tau_{\bar{K}(rr')}^{I_{\bar{K}}}
\mathcal{K}_{\bar{K},r';\bar{K},n}^{I_{\bar{K}}I_{\bar{K}}} \\
& \hspace{1.5cm}+\sum_{rr'}\mathcal{M}_{d,m;N,r}^{I_{d}I_{N}}\tau_{N(rr')}^{I_{N}}
\mathcal{K}_{N,r';\bar{K},n}^{I_{N}I_{\bar{K}}}.
\end{split}
\label{eq.5}
\end{equation}

Here, the operators $\mathcal{K}_{i,m;j,n}^{I_{i}I_{j}}$ are the three-body transition 
amplitudes, which describe the dynamics of the three-body $\bar{K}Nd$ system. To define 
the spectator particles or interacting particles in each subsystem, we used the $i, j$ 
and $k$ indices and the isospin of the interacting particles is defined by $I_ {i}$. Since 
some potentials are rank-$2$, the indices $n,l,m$ are used to define, which term of the 
sub-amplitudes is used. The operators $\mathcal{M}_{i,m;j,n}^{I_{i}I_{j}}$ are Born terms, 
which describe the effective particle-exchange potential realized by the exchanged particle 
between the quasi-particles in channels $i$ and $j$. The Born terms are defined by
\begin{equation}
\begin{split}
& \mathcal{M}_{i,m;j,n}^{I_{i}I_{j}}(p_{i},p_{j};z)= 
\frac{\Omega_{I_{i}I_{j}}}{2} \\ 
& \hspace{2.5cm}\times\int_{-1}^{+1}\, dx\, 
\frac{g^{I_{i}}_{i,m}(q_{i})\, g^{I_{j}}_{j,n}(q_{j})}{z-\frac{p_{i}^{2}}{2m_{i}}-
\frac{p_{j}^{2}}{2m_{j}}-\frac{(\vec{p}_{i}+\vec{p}_{j})^{2}}{2m_{k}}},
\end{split}
\label{eq.6}
\end{equation}
where the parameters $\Omega_{I_{i}I_{j}}$ are the spin and isospin coupling coefficients. 
The momenta $\vec{q}_{i}(\vec{p}_{i},\vec{p}_{j})$ and $\vec{q}_{j}(\vec{p}_{i},\vec{p}_{j})$ 
are given in terms of $\vec{p}_{i}$ and $\vec{p}_{j}$. We use the relations
\begin{equation}
\begin{split}
& \vec{q}_{i}=-\vec{p}_{j}-\frac{m_{j}}{m_{j}+m_{k}}\vec{p}_{i}, \\
& \vec{q}_{j}=\vec{p}_{i}+\frac{m_{i}}{m_{i}+m_{k}}\vec{p}_{j},
\end{split}
\label{eq.7}
\end{equation}
where $m_{k}$ is exchanged particle or quasi-particle mass and $x$ is defined by 
$x=\hat{p}_{i}\cdot\hat{p}_{j}$.

For the $K^{-}+\,^{3}\mathrm{He}$ reaction, the initial state in the laboratory frame 
contains an incoming kaon, the projectile, and one $^{3}\mathrm{He}$, the target, at 
rest. There are three possibilities for the final state. In one, the $^{3}\mathrm{He}$ 
survives scattering, i.e. the final state contains a kaon and a $^{3}\mathrm{He}$. This 
is called elastic scattering. The other is where the system goes to $(\bar{K}N)+d$ and 
$(\bar{K}d)+N$ channel. \\
\section{Results and discussion}
\label{result}
Before we proceed to discus about the obtained results, we shall begin with a survey on the two-body 
interactions, which are the central input to our present three-body calculations. For all two-body 
interactions, the angular momentum is taken to be zero and all potentials have the separable form in 
momentum representation.

Different phenomenological and chiral based potentials were used to describe the 
$\bar{K}N-\pi\Sigma$ interaction, which is the most important input in the $\bar{K}Nd-\pi\Sigma{d}$ 
three-body system. The phenomenological potentials SIDD$^{1}$ and SIDD$^{2}$ from Ref.~\cite{sh4} 
are constructed to reproduce the SIDDHARTA~\cite{sidd} experiment results. SIDD$^{1}$ and SIDD$^{2}$ 
potentials reproduce the one- and two-pole structure of $\Lambda(1405)$, respectively. The potentials 
have the following form
\begin{equation}
V^{I}_{\alpha\beta}(k,k')=g^{I}_{\alpha}(k)\, \lambda^{I}_{\alpha\beta} 
\, g^{I}_{\beta}(k').
\label{eq.11}
\end{equation}

Here, the strength parameters and the form factors of the two-body potential are labeled by particle 	
indices $\alpha$ and $\beta$ to take into account the coupling between $\bar{K}N$ and $\pi\Sigma$ 
systems. To study the dependence of the results to the $\bar{K}N-\pi\Sigma$ model of interaction, we 
also used the potential given by Akaishi and Yamazaki~\cite{ak1} and the new potential given in Ref.~\cite{rev} 
which the first one is an extremely deep potential. We referred to these potentials as AY and Rev-A 
potentials, respectively. The Rev-A model is a chiral based but energy-independent potential which 
is a rather new and probably not well known. Five different interaction models (A-E) are presented 
in Ref.~\cite{rev}. As the A model reproduce the lowest value of $\chi$, the A version was chosen 
to be used in present calculations. 
Most of chiral potentials in the literatures are not suitable for Faddeev calculation or at least will 
make the calculations difficult. The last $\bar{K}N$ potential that we used in our calculations is an 
energy-dependent chiral potential (Chiral-IKS). The parameters of the energy-dependent chiral based 
potential are presented in Ref.~\cite{ik3}. In Table~\ref{ta.1}, the pole position(s) of the $\bar{K}N$ 
system for all models of interaction are presented.
\begin{table*}[t!]
\caption{The pole position(s) (in MeV) of the $\bar{K}N$ system for different 
phenomenological and chiral based models of the $\bar{K}N-\pi\Sigma$ interaction. 
$\mathrm{X}^{1}$ and $\mathrm{X}^{2}$ stands for a one- and a two-pole version 
of the $\mathrm{SIDD}$ potentials. The quantities $\mathrm{p_{1}}$ and $\mathrm{p_{2}}$ 
represent the first and second pole of $\bar{K}N-\pi\Sigma$ system for each potential.}
\centering
\begin{tabular}{cccccc}
\hline\hline\noalign{\smallskip}\noalign{\smallskip}
&\,  $\mathrm{SIDD}^{1}$ \, & \, $\mathrm{SIDD}^{2}$ \, & \, $\mathrm{Chiral-IKS}$ &\,  
$\mathrm{AY}$ \, & \, $\mathrm{Rev}-A$ \, \\
\noalign{\smallskip}\noalign{\smallskip}\hline
\noalign{\smallskip}\noalign{\smallskip}
$\mathrm{p_{1}}$ \, & \, $1428.1-i46.6$ \, & \, $1418.1-i56.9$ \, & \, $1420.6-i20.3$ 
 & \, $1407.6-i20.0$ \, & \, $1422.0-i20.0$ \, \\
$\mathrm{p_{2}}$ \, & \, $-$ \, & \, $1382.0-i104.2$ \, & \, $1343.0-i72.5$ 
& \, $-$ \, & \, $-$ \, \\
\noalign{\smallskip}\noalign{\smallskip}
\hline\hline
\end{tabular}
\label{ta.1} 
\end{table*}

Two models of interaction were used to describe the $pd$ interaction. The first one is a 
two terms potential, which includes the short range repulsive part of the interaction
\begin{equation}
V^{Nd}_{A}(k,k')=\sum_{m=1}^{2}g_{A;m}^{Nd}(k)\, \lambda_{A;m}^{Nd}\, g^{Nd}_{A;m}(k'),
\label{eq.12}
\end{equation}
where the functions $g_{i}^{Nd}(k)$ are the form factors of $pd$ interaction and are 
parametrized by Yamaguchi form~\cite{yam}:
\begin{equation}
g_{A;m}^{Nd}(k)=\frac{1}{k^2+(\Lambda_{A;m}^{Nd})^2}.
\label{eq.13}
\end{equation}

We refer to the two-term potential as $V^{Nd}_{A}$. The parameters of the $V^{Nd}_{A}$ potential 
are adjusted to reproduce the $p-d$ phase-shifts~\cite{chen}. The physical values for data fitting 
were obtained by solving the Lippmann-Schwinger equations without inclusion of the Coulomb interaction 
into the $p-d$ system. We used also a one-term potential for $pd$ interaction and its parameter are 
adjusted to reproduce the $^{3}\mathrm{He}$ binding energy and $pd$ scattering length. We refer to 
the rank one potential as $V^{Nd}_{B}$. The parameters of the the $V^{Nd}_{A}$ and $V^{Nd}_{B}$ 
potentials are presented in Table~\ref{ta.2}
\begin{table}[H]
\caption{The parameters of $V^{Nd}_{A}$ and $V^{Nd}_{B}$ potentials to describe 
the $pd$ interaction. The range parameters are in MeV and the strength parameters 
are in fm$^{-2}$.}
\centering
\begin{tabular}{ccccc}
\hline\hline\noalign{\smallskip}\noalign{\smallskip}
$V^{Nd}_{A}$ potential: & & & & \\
\noalign{\smallskip}
& $\Lambda_{A;1}^{Nd}$ \, & $\Lambda_{A;2}^{Nd}$\, & $\lambda_{A;1}^{Nd}$\, & $\lambda_{A;2}^{Nd}$ \\
\noalign{\smallskip}
& 115\, & 152\, & -0.0404\, & 0.2967 \\
\noalign{\smallskip}\hline\hline\noalign{\smallskip}
$V^{Nd}_{B}$ potential: & & & & \\
\noalign{\smallskip}
& $\Lambda_{B}^{Nd}$ \, & $\lambda_{B}^{Nd}$\, & &  \\
\noalign{\smallskip}
& 139.1\, & -0.0037\, &  &  \\
\noalign{\smallskip}
\hline\hline
\end{tabular}
\label{ta.2} 
\end{table}

We need also a potential model to describe the interaction between antikaon and deuteron. 
A one-channel complex potential with rank-$2$ were used to describe $K^{-}d$ interaction
\begin{equation}
V^{\bar{K}d}(k,k')=\sum_{m=1}^{2}g^{\bar{K}d}_{m}(k)\, \lambda^{\bar{K}d}_{m;Complex}\, 
g^{\bar{K}d}_{m}(k').
\label{eq.14}
\end{equation}

The parameters of this potential are given in Ref.~\cite{sh5}. The complex strength parameters 
and range parameters of the potential are adjusted to reproduce the $K^{-}d$ scattering length 
$a_{K^{-}d}$ and also the effective range $r^{0}_{K^{-}d}$~\cite{sh5}.

\begin{figure*}[ht]
\centering
\includegraphics[width=8.5cm]{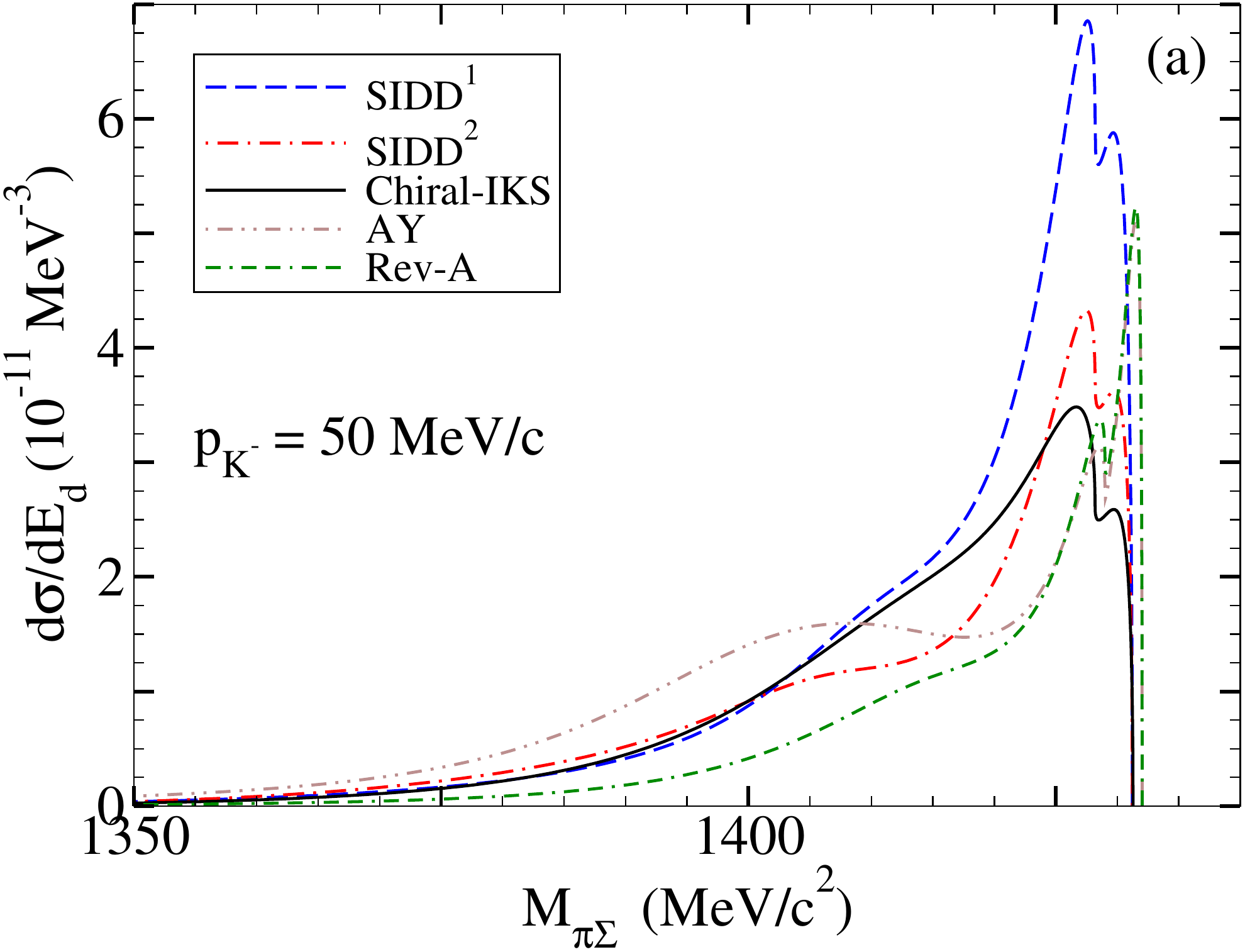} 
\hspace{0.5cm}
\includegraphics[width=8.5cm]{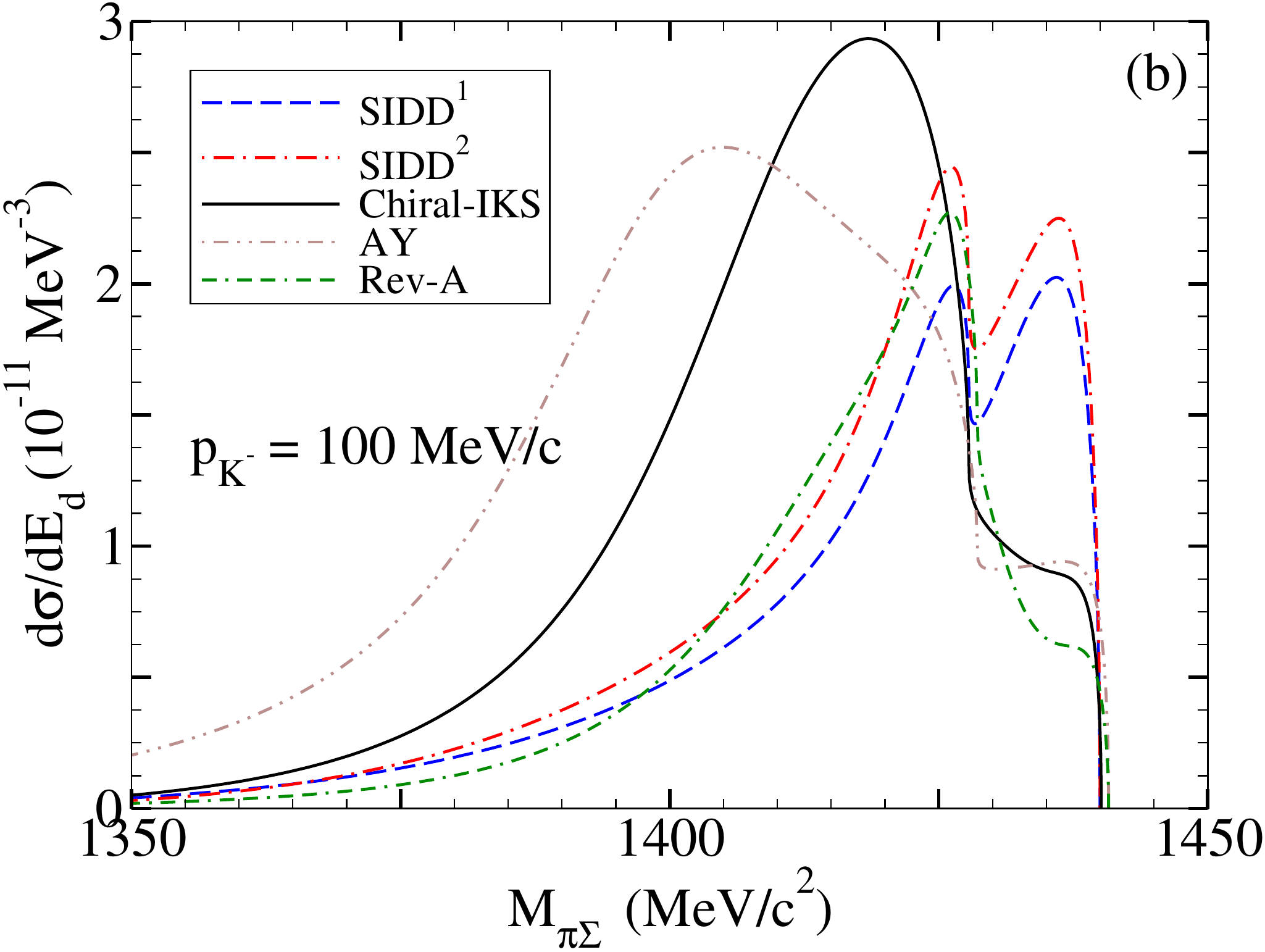} \\
\includegraphics[width=8.5cm]{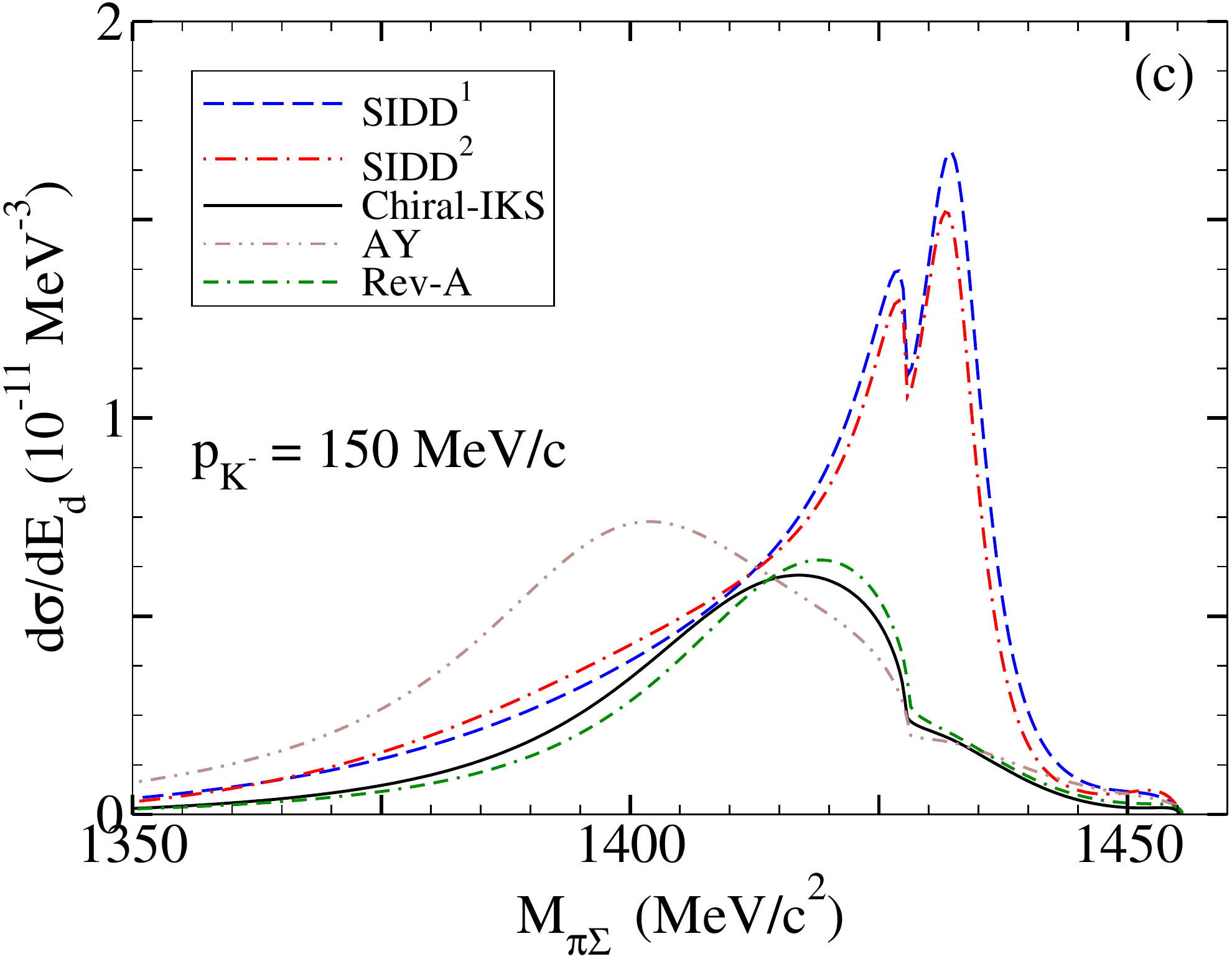}
\hspace{0.5cm}
\includegraphics[width=8.5cm]{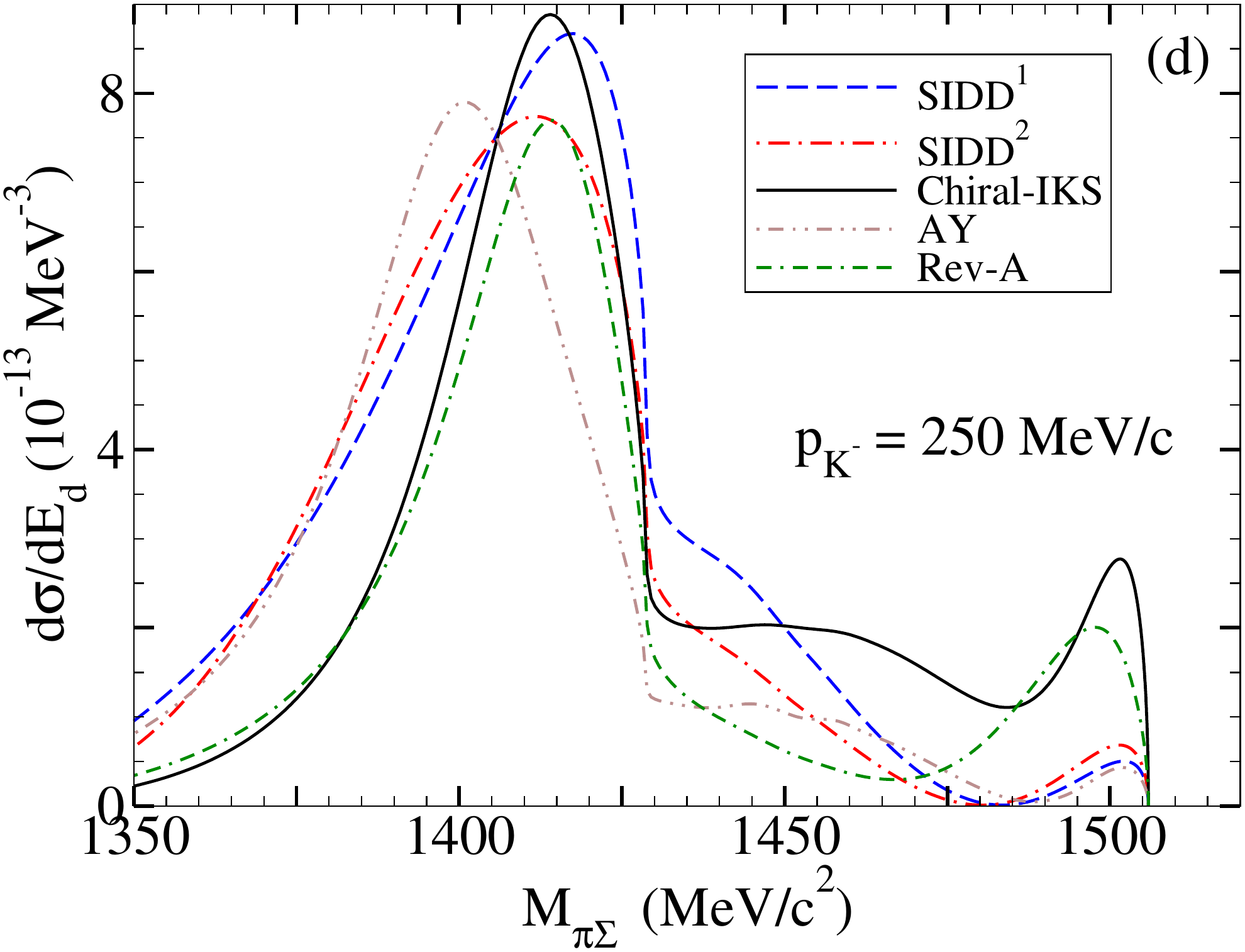}
\caption{(Color on line) The $\pi\Sigma$ mass spectra for $K^{-}+\,^{3}\mathrm{He}\rightarrow\pi^{0}\Sigma^{0}+d$ 
reaction. Different types of $\bar{K}N-\pi\Sigma$ potentials were used. The kaon incident 
momentum is around $p_{\bar{K}}^{cm}=50-250$ MeV/c. In panels (a), (b), (c) and (d), the 
values of $\bar{P}_{K}$ are 50, 100, 150 and 250 MeV/c, respectively. The blue dashed and 
red dash-dotted curves show the mass spectrum with $\mathrm{SIDD}^{1}$ and $\mathrm{SIDD}^{2}$ 
potential, respectively. The extracted results for Chiral-IKS, AY and Rev-A potentials are 
also depicted by black solid, brown dash-dot-dotted and green dash-dash-dotted lines, 
respectively.}
\label{fig.3}
\end{figure*}
\subsection{The one-channel AGS calculation of the $\bar{K}Nd-\pi\Sigma{d}$}
\label{dir}
To take the coupling between $\bar{K}N$ and $\pi\Sigma$ channels into account, the formalism of 
Faddeev equations should be extended to include the $\pi\Sigma{d}$ channel. In the present subsection, 
the $\pi\Sigma{d}$ channel of the $\bar{K}Nd$ system has not been included directly and one-channel 
Faddeev AGS equations are solved for three-body $\bar{K}Nd$ system and we approximated the full 
coupled-channel interaction by using the so-called exact optical $\bar{K}N(-\pi\Sigma)$ potential~\cite{sh3}. 
Therefore, the decaying to the $\pi\Sigma{d}$ channel is taken into account through the imaginary 
part of the optical $\bar{K}N(-\pi\Sigma)$ potential. To study the possible signature of the 
$\Lambda(1405)$ resonance in the $\pi^{0}\Sigma^{0}$ mass spectra in the $K^{-}+(Nd)\rightarrow\pi\Sigma+d$ 
reaction, first we should define break-up amplitude. As we do not include the lower lying channels 
directly into the calculations, the only Faddeev amplitude, which contribute in the scattering 
amplitude is $\mathcal{K}_{d,\bar{K}}$. Therefore, the break-up amplitude can be expressed as 
\begin{equation}
\begin{split}
& T_{(\pi\Sigma)+{d}\leftarrow \bar{K}+(Nd)} (\vec{k}_{d},\vec{p}_{d},\bar{p}_{\bar{K}};z)= 
\sum_{n=1}^{2}g^{I=0}_{\pi\Sigma}(\vec{k}_{d} )\\
& \hspace{0.5cm}\times \tau^{I=0}_{{(\pi\Sigma)}\leftarrow(\bar{K}N)}
(z-\frac{p^{2}_{d}}{2\eta_{d}}) 
\mathcal{K}^{0\frac{1}{2}}_{d,1;\bar{K},n}(p_{d},\bar{p}_{\bar{K}};z),
\end{split}
\label{eq.8}
\end{equation}
where $\vec{k}_{i}$ is the relative momentum between the interacting pair ($jk$) and 
$\bar{p}_{\bar{K}}$ is the initial momentum of $\bar{K}$ in $\bar{K}Nd$ center of mass. 
The quantity $\mathcal{K}^{I_{i}I_{j}}_{i(m),j(n)}$ is the Faddeev amplitude, which is 
derived from Faddeev equation (\ref{eq.5}). 

Using Eq.(\ref{eq.8}), we define the break-up cross section of 
$\bar{K}+(Nd)\rightarrow{d}+(\pi^{0}\Sigma^{0})$ as follows:
\begin{equation}
\begin{split}
& \frac{d\sigma}{dE_{d}}=\frac{\omega_{(Nd)} \omega_{\bar{K}}}{z\bar{p}_{\bar{K}}}
\frac{m_{\pi}m_{\Sigma}m_{d}}{m_{\pi}+m_{\Sigma}+m_{d}}
\int{d}\Omega_{p_{d}}d\Omega_{k_{d}}p_{d}k_{d} \\
& \hspace{0.7cm}\times\sum_{if}
|T_{(\pi\Sigma)+{d}\leftarrow \bar{K}+(Nd)} (\vec{k}_{d},\vec{p}_{d},\bar{p}_{\bar{K}};z)|^2,
\end{split}
\label{eq.9}
\end{equation}
where $E_{d}$ is the deuteron energy in the center-of-mass frame of $\pi\Sigma$, 
which is defined by
\begin{equation}
E_{d}=m_{d}+\frac{p^{2}_{d}}{2\eta_{d}},
\label{eq.10}
\end{equation}
and the energies $\omega_{(Nd)}$ and $\omega_{\bar{K}}$ are the kinetic energy of 
${\bar{K}}$ and $Nd$ in the initial state.\\

Since the input energy of the AGS equations is above the $\bar{K}Nd$ threshold, the moving 
singularities will appear in the three-body amplitudes. To remove these standard singularities, 
we have followed the same procedure implemented in Refs.~\cite{poin1,poin2}. Using the so called 
\textquotedblleft{point-method}\textquotedblright, we computed the cross section of 
$K^{-}+(Nd)\rightarrow\pi\Sigma+d$ reaction.

Starting from Faddeev AGS equations~\ref{eq.5}, the $\pi^{0}\Sigma^{0}$ invariant mass for 
$K^{-}+(Nd)\rightarrow\pi\Sigma+d$ reaction was calculated. In our calculation, we studied 
the dependence of the mass spectrum on the fundamental $\bar{K}N-\pi\Sigma$ interaction by 
using five different interaction models reproducing various pole structure for $\Lambda(1405)$ 
resonance. With this method, we extracted the $\pi\Sigma$ mass spectrum for different incident 
antikaon momentum $p_{\bar{K}}^{cm}=50-250\mathrm{MeV}/c$. The extracted mass spectrum for 
momenta $p_{\bar{K}}^{cm}=50-150\mathrm{MeV}/c$ is strongly affected by threshold effects, 
but for the momentum $p_{\bar{K}}^{cm}=250\mathrm{MeV}/c$ the signature of the resonance is 
clearly visible. Furthermore, it was found that the $\pi\Sigma$ mass spectrum, reflecting the 
$\Lambda$(1405) mass distribution and width, depends quite sensitively on the $\bar{K}N-\pi\Sigma$ 
model of interaction.

A definitive study of the three-body $\bar{K}Nd$ system could be also performed using the 
standard energy-dependent $\bar{K}N$ input potential, too~\cite{ik3}. The energy-dependent 
potentials provide a weaker $\bar{K}N$ attraction for lower energies than the energy-independent 
potentials. Therefore, the quasi-bound state in $\bar{K}N$ system resulting from the 
energy-dependent potential happens to be shallower. In Fig.~\ref{fig.3}, a comparison is 
made between the results obtained for the chiral-IKS $\bar{K}N-\pi\Sigma$ and the calculated 
mass spectra for other potentials. For energy-dependent potential the peak structure is 
not located at the position of the first pole given in table~\ref{ta.1}. These results 
are in agreement with the statement that the $\Lambda(1405)$ spectrum is the superposition 
of two independent states and one can not see two different pole structure in the 
$\Lambda(1405)$ spectrum~\cite{os1,ol1}.

The key point in the present calculations is the detection of deuteron which is a loosely bound 
system. Although, for large momenta the signal of $\Lambda$(1405) overcomes the kinetic effects, 
at these energies, the probability for the deuteron (as a loosely bound system) to survive the 
reaction will decrease and one would expect a rather low reaction rate for 
$K^{-}+\, ^{3}\mathrm{He}\rightarrow\pi\Sigma{d}$ reaction. Therefore, an accurate measurements 
of the $\pi\Sigma$ mass distribution at an optimized value of momentum in a possible future 
experiment could give a good reaction rate and less kinematical effects.

\begin{figure}[H]
\centering
\includegraphics[width=8.cm]{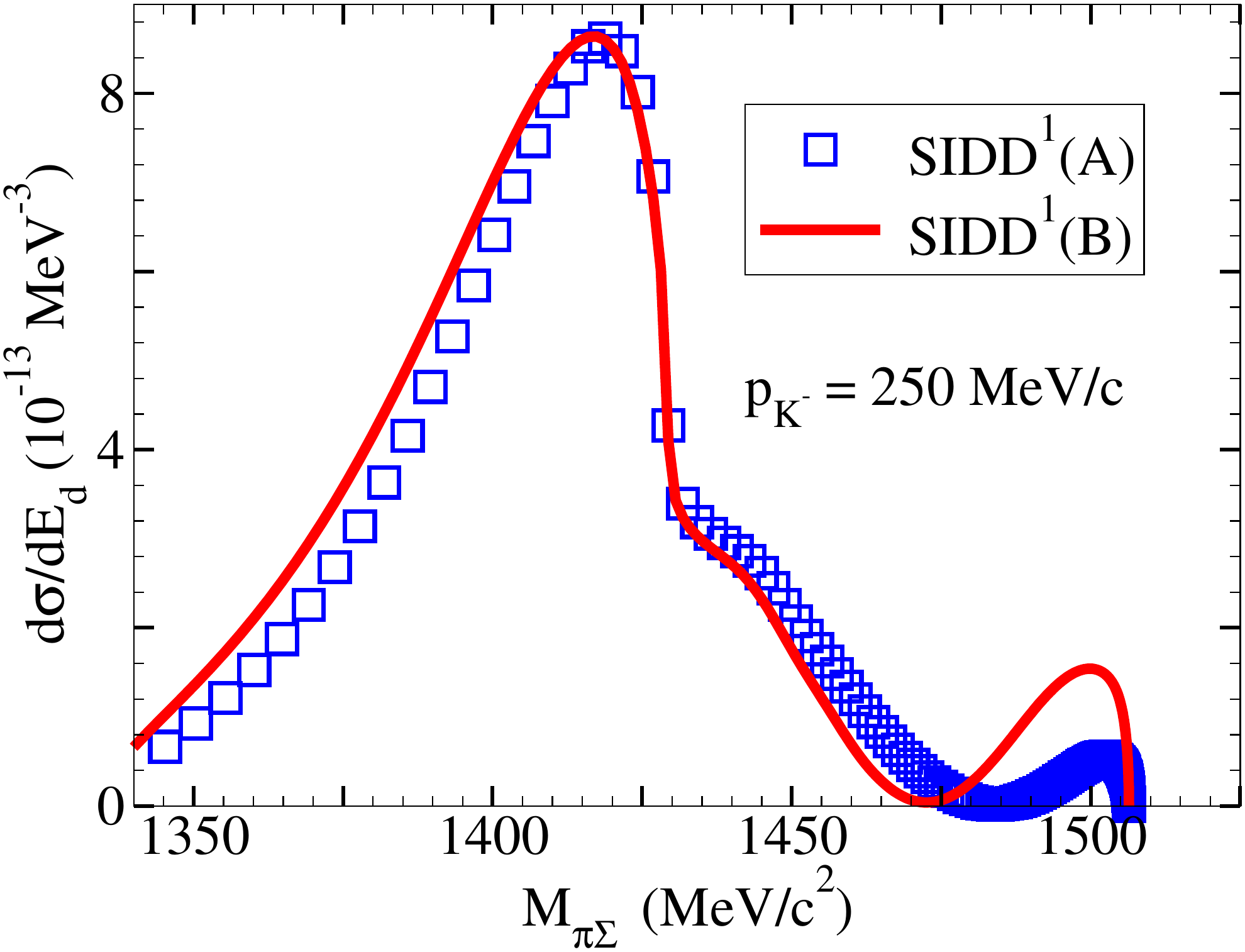}
\caption{(Color online) The invariant mass spectra for $K^{-}+\,^{3}\mathrm{He}$ 
reaction. We used the one-term (solid curves) and two-term (square symbols) potential 
for $Nd$ interaction to study the dependence of the $\pi\Sigma$ mass spectra to the 
$Nd$ model of interaction. The symbols $A$ and $B$ are corresponding to $V^{Nd}_{A}$ 
and $V^{Nd}_{B}$, respectively. In our calculations, the one-pole version of SIDD 
potential was used to describe the $\bar{K}N$ interaction.}
\label{fig.4}
\end{figure}

To study the dependence of the $\pi\Sigma$ invariant mass on the $Nd$ model of interaction, 
in Fig.~\ref{fig.4}, we calculated the $\pi\Sigma$ mass spectra for two different $Nd$ 
interaction, including a one-term ($V^{Nd}_{B}$ ) and a two-term ($V^{Nd}_{A}$) potential. 
Comparing the extracted invariant mass spectra for one-term potential and the corresponding 
mass spectra for the two-term potential, we can see that the $Nd$ interaction can affect 
the mass spectra especially, for energies above the $\bar{K}+Nd$ mass threshold. However, 
the mass spectrum in energy region around the $\bar{K}N$ pole position did not change 
seriously by changing the $Nd$ model of interaction.
\subsection{The coupled-channel AGS calculation of the $\bar{K}Nd-\pi\Sigma{d}$}
\label{dev}
In subsection~\ref{dir}, we solved the one-channel AGS equations for $\bar{K}Nd$ system and the decaying to 
the lower lying channels is included by using the so-called exact optical potential for $\bar{K}N$ interaction. 
In one-channel Faddeev calculations the effect of the $\tau_{\pi\Sigma\rightarrow\pi\Sigma}$ amplitude was 
excluded. Based on chiral unitary approach, the first and second pole of $\Lambda$(1405) have clearly different 
coupling nature to the meson-baryon channels; the higher energy pole dominantly couples to the $\bar{K}N$ 
channel, while the lower energy pole strongly couples to the $\pi\Sigma$ channel. Due to the different coupling 
nature of these resonances, the shape of the $\Lambda$(1405) spectrum can be different depending on the initial 
and final channels. In the $\bar{K}N\rightarrow\pi\Sigma$ amplitude, the initial $\bar{K}N$ channel gets more 
contribution from the higher pole with a larger weight. Consequently, the spectrum shape has a peak around 1420 
MeV coming from the higher pole~\cite{os1,ol1}. This is obviously different from the $\pi\Sigma\rightarrow\pi\Sigma$ 
spectrum which is largely affected by the lower pole. Therefore, the extracted mass spectra in subsection~\ref{dir} 
can not reproduce exactly the possible experimental mass spectra. To calculate the exact $\pi\Sigma$ mass spectra 
for $\bar{K}+(Nd)$ reaction, we solved the Faddeev equations for coupled-channel $\bar{K}Nd-\pi\Sigma{d}$ system. 
In addition to the above mentioned reaction, we need an interaction model for $\Sigma{d}$ and $\pi{d}$ subsystems. 
In present calculations, the effect of $\pi{d}$ interaction is neglected. To describe the $\Sigma{d}$ interaction, 
we used a one term complex potential in a form given in Eq.~\ref{eq.14}. To define the parameters of the $\Sigma{d}$ 
interaction, we used the $\Sigma{d}$ scattering length, which can be extracted from the Faddeev equations of the 
three-body $\Sigma{(NN)_{s=1,I=0}}$ system. The antisymmetric Faddeev equations for $\Sigma{d}$ system can be given by

\begin{equation}
\begin{split}
& \mathcal{K}_{\Sigma,\Sigma}^{sI,s'I'}=\sum_{s''I''}
\mathcal{M}_{\Sigma,N_{1}}^{sI,s''I''}
\,\tau_{N_{1}}^{s''I''}\,(\mathcal{K}_{N_{1},\Sigma}^{s''I'',s'I'}-
\mathcal{K}_{N_{2},\Sigma}^{s''I'',s'I'}) \\
& \mathcal{K}_{N_{1},\Sigma}^{sI,s'I'}-\mathcal{K}_{N_{2},\Sigma}^{sI,s'I'} 
=2\mathcal{M}_{N_{1},\Sigma}^{sI,s'I'} \\
& \hspace{1cm}+\sum_{s''I''} 2\mathcal{M}_{N_{1},\Sigma}^{sI,s''I''}\,
\tau_{\Sigma}^{s''I''}\,\mathcal{K}_{\Sigma,\Sigma}^{s''I'',s'I'} \\
& \hspace{1cm}-\sum_{s''I''} \mathcal{M}_{N_{1},N_{2}}^{sI,s''I''}\,\tau_{N_{2}}^{s''I''}\,
(\mathcal{K}_{N_{1},\Sigma}^{s''I'',s'I'}-
\mathcal{K}_{N_{2},\Sigma}^{s''I'',s'I'}) \\
\end{split}
\label{eq.15}
\end{equation}
where the $\Sigma{d}$ scattering length is given by
\begin{equation}
a_{\Sigma{d}}=-4\pi^{2}\,\mu_{\Sigma{d}}\,
\mathcal{K}^{\frac{1}{2}1,\frac{1}{2}1}_{\Sigma,\Sigma}
(p\rightarrow{0},p'\rightarrow{0};z=-E_{d})
\label{eq.16}
\end{equation}
where $\mu_{\Sigma{d}}$ is the reduced mass of $\Sigma{d}$ and $E_{d}$ is the binding energy of deuteron. 
To solve Eq.~\ref{eq.15}, we need as input a potential model for $\Sigma{N}-\Lambda{N}$ and $NN$ interactions. 
In our three-body study, to describe the singlet and triplet $\Sigma{N}$ interaction, we used the potentials 
given in Ref.~\cite{sh3} and for triplet $NN$ interaction, we choose a potential of PEST type~\cite{zan}, 
which is a separablization of the Paris potential. The $\Sigma{d}$ scattering length value obtained with 
the two above mentioned $\Sigma{N}-\Lambda{N}$ and $NN$ potentials is
\begin{equation}
a_{\Sigma{d}}=-1.59+i0.71 \, \mathrm{fm}^{-1}
\label{eq.17}
\end{equation}

The whole dynamics of coupled-channel $\bar{K}Nd-\pi\Sigma{d}$ three-body system is described in terms of 
the transition amplitudes $\mathcal{K}_{i,m;j,n}^{\alpha\beta;I_{i}I_{j}}$. The superscripts $\alpha,\beta=1,2$ 
are included to take into account the coupling between $\bar{K}Nd$ and $\pi\Sigma{d}$ systems. The Faddeev 
AGS equations for $\bar{K}Nd-\pi\Sigma{d}$ system can be expressed by
\begin{equation}
\begin{split}
& \mathcal{K}_{\bar{K},m;\bar{K},n}^{11;I_{\bar{K}}I_{\bar{K}}}= 
 \sum_{rr'}(\mathcal{M}_{\bar{K},m;N,r}^{1,I_{\bar{K}}I_{N}}\tau_{N(rr')}^{11;I_{N}}
\mathcal{K}_{N,r';\bar{K},n}^{11;I_{N}I_{\bar{K}}} \\
& +\mathcal{M}_{\bar{K},m;d,r}^{1,I_{\bar{K}}I_{d}}
\tau_{d(rr')}^{11,I_{d}}\mathcal{K}_{d,r';\bar{K},n}^{11,I_{d}I_{\bar{K}}}
+\mathcal{M}_{\bar{K},m;d,r}^{1,I_{\bar{K}}I_{d}}
\tau_{d(rr')}^{12,I_{d}}\mathcal{K}_{d,r';\bar{K},n}^{21,I_{d}I_{\bar{K}}}) \\
& \mathcal{K}_{N,m;\bar{K},n}^{11;I_{N}I_{\bar{K}}}= 
\mathcal{M}_{N,m;\bar{K},n}^{1;I_{N}I_{\bar{K}}}+ 
 \sum_{rr'}(\mathcal{M}_{N,m;\bar{K},r}^{1;I_{N}I_{\bar{K}}}
\tau_{\bar{K}(rr')}^{11,I_{\bar{K}}}
\mathcal{K}_{\bar{K},r';\bar{K},n}^{11,I_{\bar{K}}I_{\bar{K}}} \\
& +\mathcal{M}_{N,m;d,r}^{1,I_{N}I_{d}}\tau_{d(rr')}^{11,I_{d}}
\mathcal{K}_{d,r';\bar{K},n}^{11,I_{d}I_{\bar{K}}}
+\mathcal{M}_{N,m;d,r}^{1,I_{N}I_{d}}\tau_{d(rr')}^{12,I_{d}}
\mathcal{K}_{d,r';\bar{K},n}^{21;I_{d}I_{\bar{K}}}) \\
& \mathcal{K}_{d,m;\bar{K},n}^{11;I_{d}I_{\bar{K}}} = 
\mathcal{M}_{d,m;\bar{K},n}^{1;I_{d}I_{\bar{K}}}+
 \sum_{rr'}(\mathcal{M}_{d,m;\bar{K},r}^{1;I_{d}I_{\bar{K}}}
\tau_{\bar{K}(rr')}^{11;I_{\bar{K}}}
\mathcal{K}_{\bar{K},r';\bar{K},n}^{11;I_{\bar{K}}I_{\bar{K}}} \\
& \hspace{1.4cm}+\mathcal{M}_{d,m;N,r}^{1;I_{d}I_{N}}\tau_{N(rr')}^{11;I_{N}}
\mathcal{K}_{N,r';\bar{K},n}^{11;I_{N}I_{\bar{K}}}) \\
& \mathcal{K}_{d,m;\bar{K},n}^{21;I_{d}I_{\bar{K}}} = 
\sum_{rr'}\mathcal{M}_{d,m;\pi,r}^{2;I_{d}I_{\pi}}
\tau_{\pi(rr')}^{22;I_{\pi}}
\mathcal{K}_{\pi,r';\bar{K},n}^{21;I_{\pi}I_{\bar{K}}} \\
& \mathcal{K}_{\pi,m;\bar{K},n}^{21;I_{d}I_{\bar{K}}} = 
\sum_{rr'}(\mathcal{M}_{\pi,m;d,r}^{2;I_{\pi}I_{d}}
\tau_{d(rr')}^{22;I_{d}}
\mathcal{K}_{d,r';\bar{K},n}^{21;I_{d}I_{\bar{K}}} \\
& \hspace{1.4cm}+\mathcal{M}_{\pi,m;d,r}^{2;I_{\pi}I_{d}}\tau_{d(rr')}^{21;I_{d}}
\mathcal{K}_{d,r';\bar{K},n}^{11;I_{d}I_{\bar{K}}}).
\end{split}
\label{eq.18}
\end{equation}

\begin{figure*}[htb]
\centering
\includegraphics[width=8.5cm]{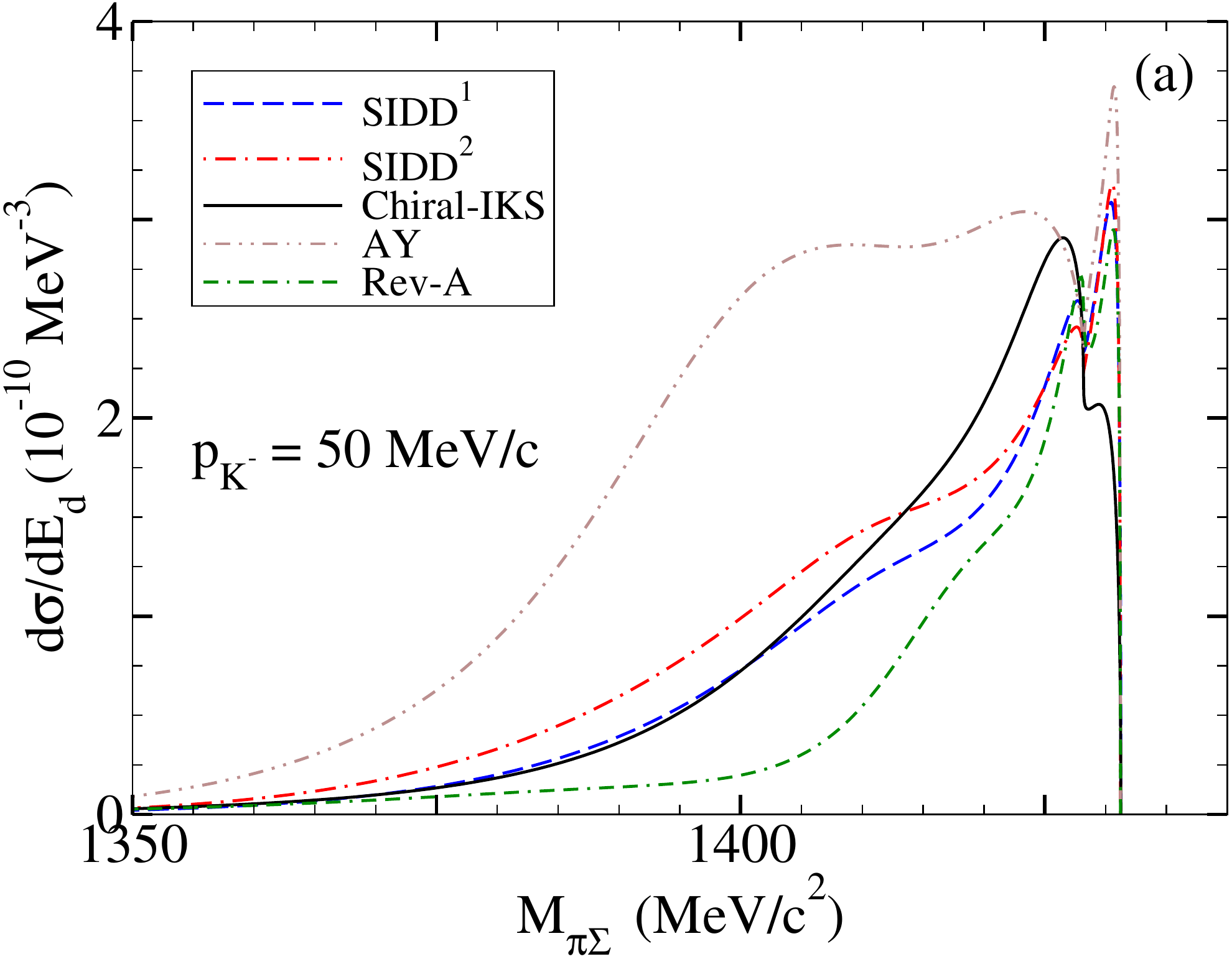} 
\hspace{0.5cm}
\includegraphics[width=8.5cm]{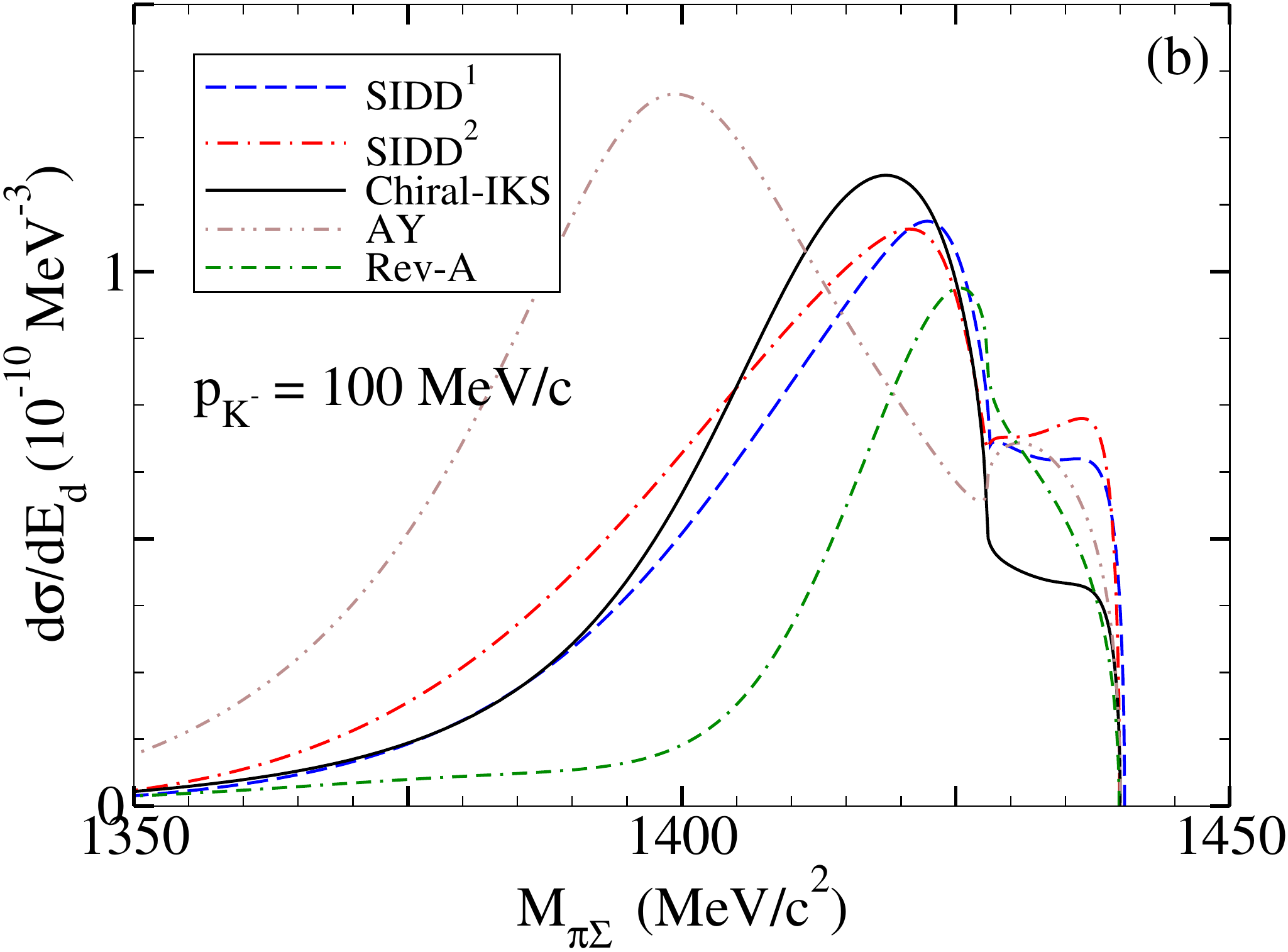} \\
\includegraphics[width=8.5cm]{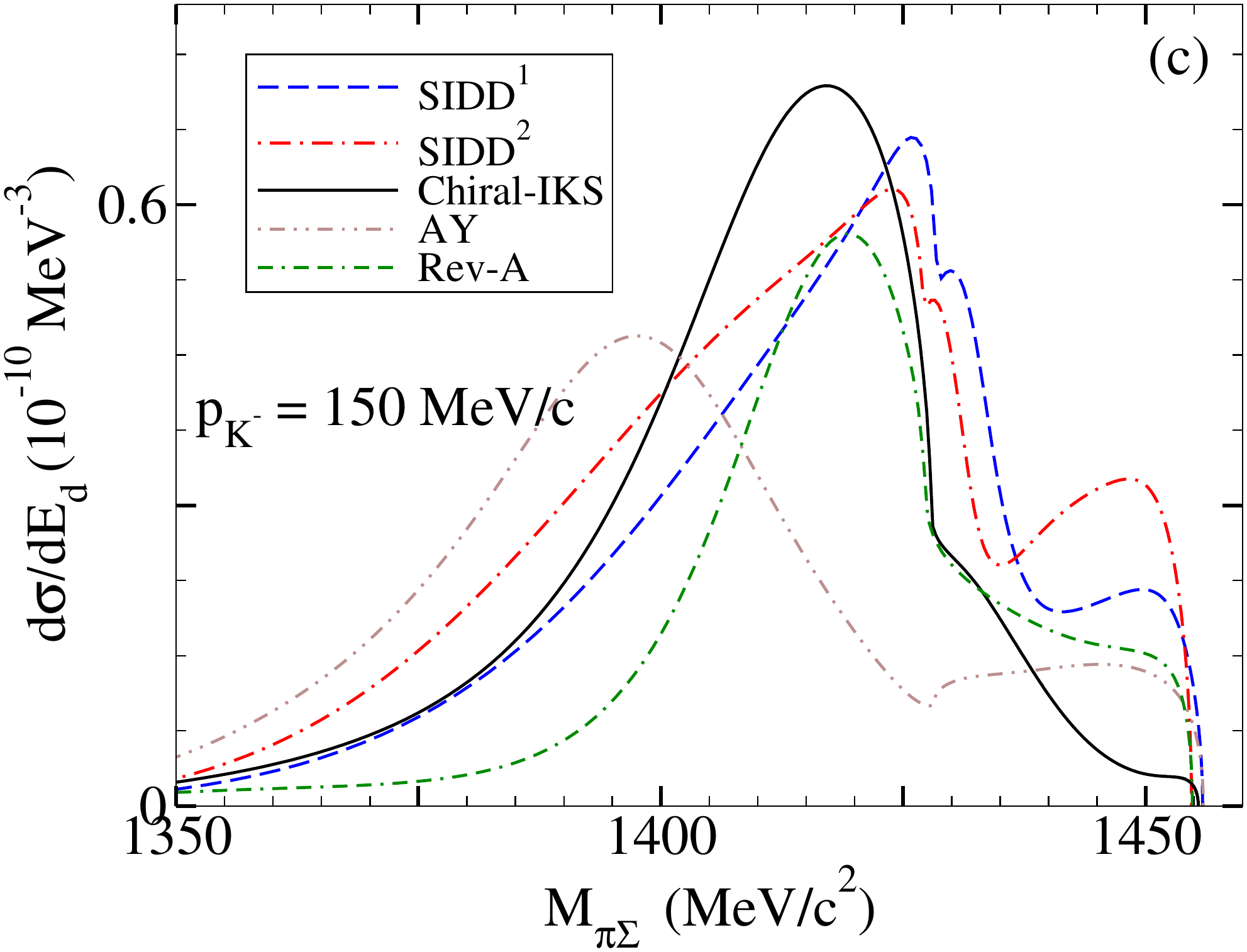}
\hspace{0.5cm}
\includegraphics[width=8.5cm]{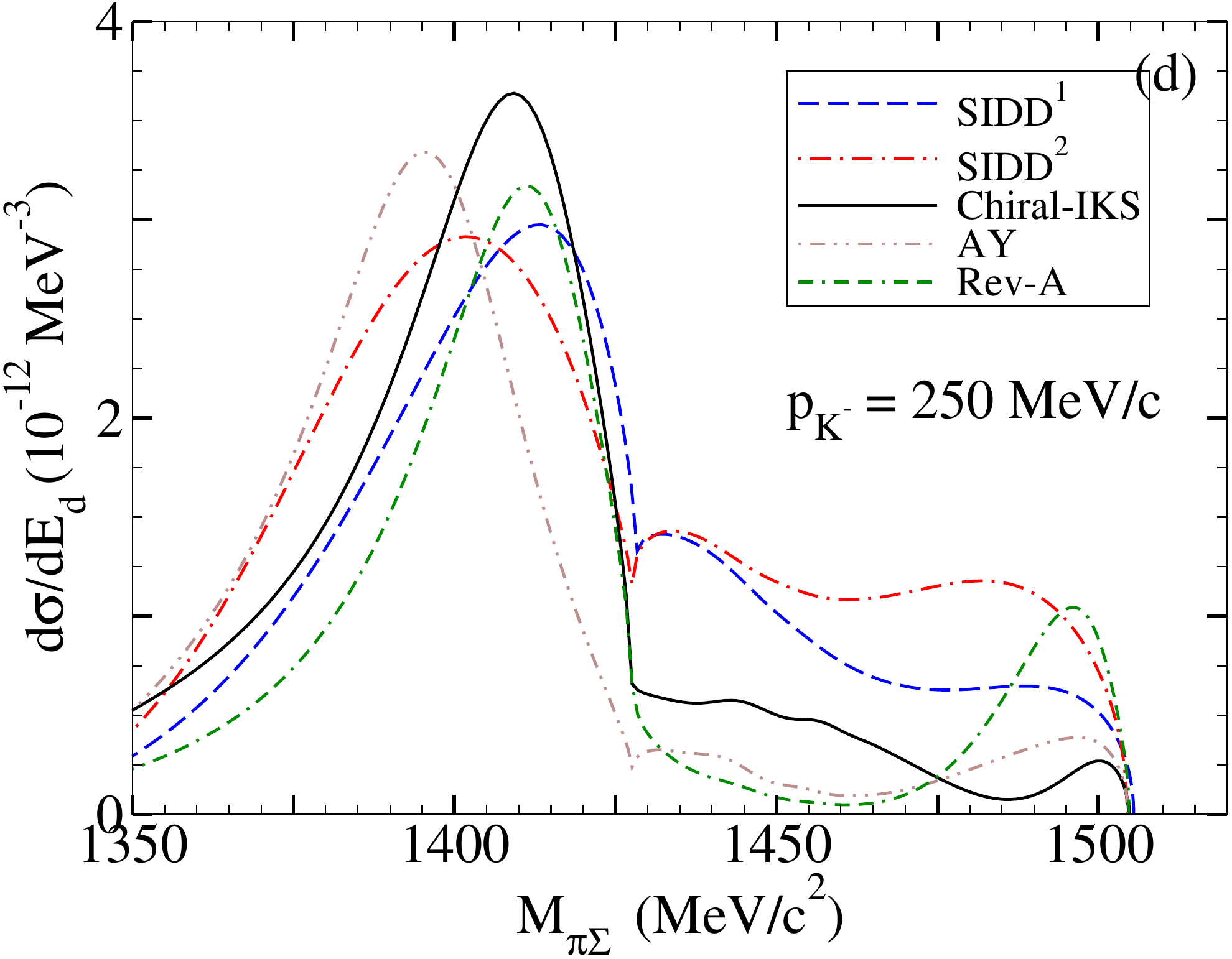}
\caption{(Color on line) Same as Fig.\ref{fig.3}, but in the present calculations, 
the full coupled-channel Faddeev AGS equations for $\bar{K}Nd-\pi\Sigma{d}$ 
system are solved.}
\label{fig.5}
\end{figure*}

The break-up amplitude for $\bar{K}+(Nd)\rightarrow (\pi\Sigma)+d$ reaction in 
terms of the Faddeev transition amplitudes can be given by
\begin{equation}
\begin{split}
& T_{(\pi\Sigma)+{d}\leftarrow(Nd)+\bar{K}} 
(\vec{k}_{d},\vec{p}_{d},\bar{P}_{\bar{K}};z) \\
& =\sum_{n}g_{\pi\Sigma}^{I=0}(\vec{k}_d )\tau_{\pi\Sigma
\leftarrow\bar{K}N}^{I=0}(z-E_{d}(\vec{p}_{d}))
\mathcal{K}_{d,1;\bar{K},n}^{11;;0\frac{1}{2}}(p_{d},\bar{P}_{\bar{K}};z) \\
& +\sum_{n}g_{\pi\Sigma}^{I=0}(\vec{k}_{d} )\tau_{\pi\Sigma
\leftarrow\pi\Sigma}^{I=0}(z-E_N(\vec{p}_d ))
\mathcal{K}_{d,1;\bar{K},n}^{21;;0\frac{1}{2}}(p_{d},\bar{P}_{\bar{K}};z) \\
& +\sum_{n}\langle[\pi\otimes\Sigma]_{I=0}\otimes{d}\mid\pi
\otimes[\Sigma\otimes{d}]_{I=1}
\rangle{g}_{\Sigma d}^{I=1}(\vec{k}_{\pi}) \\
& \hspace{1.cm}\times \tau_{\Sigma{d}\leftarrow
\Sigma{d}}^{I=1}(z-E_{\pi}(\vec{p}_{\pi}))
\mathcal{K}_{\pi,1;\bar{K},n}^{21;1\frac{1}{2}}(p_{\pi},\bar{P}_{\bar{K}};z),
\end{split}
\label{eq.19}
\end{equation}
where the momenta $\vec{p}_{\pi}$ and $\vec{k}_{\pi}$ are given by
\begin{equation}
\begin{split}
&\vec{p}_{\pi}=\vec{k}_{d}-\frac{m_{\pi}}{m_{\pi}+m_{\Sigma}}\vec{p}_{d} \\
&\vec{k}_{\pi}=-\frac{m_{d}}{m_{\Sigma}+m_{d}}\vec{k}_{d}      
-\frac{m_{\Sigma}(m_{\pi}+m_{\Sigma}+m_{d})}{(m_{\pi}+m_{\Sigma})(m_{\Sigma}+m_{d})}\vec{p}_{d}.
\end{split}
\label{eq.20}
\end{equation}

As one see from Eq.~\ref{eq.19}, in coupled-channel calculations plus the $\mathcal{K}_{d,1;\bar{K},n}^{11;;0\frac{1}{2}}$ 
amplitude, the effect of the $\mathcal{K}_{d,1;\bar{K},n}^{21;;0\frac{1}{2}}$ and $\mathcal{K}_{\pi,1;\bar{K},n}^{21;1\frac{1}{2}}$ 
are also included which accordingly, produces a more precise mass spectrum for $\pi\Sigma$. Inserting 
the new break-up amplitude (Eq.~\ref{eq.19}) in Eq.~\ref{eq.9}, we can calculate the $\pi\Sigma$ mass 
spectrum for $\bar{K}+(Nd)\rightarrow (\pi\Sigma)+d$ reaction. In Fig.~\ref{fig.5}, we calculated the 
$\pi\Sigma$ mass spectrum using different potential models for $\bar{K}N-\pi\Sigma$ interaction. As one 
can see from Fig.~\ref{fig.5}, in the fully coupled-channel calculations the resonance part of the mass 
spectrum is stronger and a more clear peak structure can be seen in $\pi\Sigma$ invariant mass. However, 
the observed peak structure of each model of $\bar{K}N-\pi\Sigma$ interaction is located at lower energies 
than those presented in Table~\ref{ta.1}, due to the momentum distribution in $p-d$ subsystem. 
By comparison of results using one-channel and coupled-channel Faddeev equations, it may be possible to 
study the effect of $\tau_{\pi\Sigma\rightarrow\pi\Sigma}$ amplitude on $\pi\Sigma$ invariant mass. As can 
be seen in Fig.~\ref{fig.5}, the extracted mass spectra in coupled-channel calculations are rather 
different from those by one-channel Faddeev calculations and it may be possible to discriminate between 
these two approaches. Therefore, the one-channel Faddeev calculations cannot be a strong tool to study 
the dynamics of $\Lambda(1405)$ resonance in $\bar{K}+(Nd)\rightarrow (\pi\Sigma)+d$ reaction.

As one can see in panel (B), an accurate measurements of the $\pi\Sigma$ mass distribution at
$p_{K^{-}}=100\mathrm{MeV/c}$ can differentiate the AY and Rev-A potentials from the others 
and studying $K^{-}+\, ^{3}\mathrm{He}$ reaction at $p_{K^{-}}=250\mathrm{MeV/c}$, one have a 
chance to discriminate between the other three potentials under the consideration. Looking at 
Fig.~\ref{fig.5} one can clearly that for chiral energy-dependent potential, the magnitude of 
the mass spectrum above the $\bar{K}N$ threshold is considerably smaller than those by other 
potentials for all kaon incident momenta. Therefore, such a combined study at two different 
initial energies shows a big potential to discriminate between possible mechanisms of the 
formation of $\Lambda(1405)$ resonance.
\section{Conclusion}
\label{conc}
In summary, the Faddeev-type calculations of $\bar{K}Nd$ system with quantum numbers $I=0$ and 
$s=\frac{1}{2}$ were performed. Solving the one-channel and full coupled-channel Faddeev equations 
for $\bar{K}Nd-\pi\Sigma{d}$ system, we calculated the $\pi\Sigma$ mass spectrum resulting from 
$K^{-}+\,^{3}\mathrm{He}\rightarrow\pi\Sigma{d}$ reaction by using the deuteron mass spectrum. 
The logarithmic singularities that appear when solving the AGS equations for the real scattering 
energies have been successfully handled by making use of the point method. To investigate the 
dependence of the resulting mass spectrum on models of $\bar{K}N-\pi\Sigma$ interaction, different 
phenomenological and chiral based potentials having the one- and two-pole structure of $\Lambda(1405)$ 
resonance, were used. We have examined how well the signature of the $\Lambda(1405)$ resonance 
manifests itself in the $\pi\Sigma$ invariant mass. By comparison of results using different 
interaction models, it was found that it may be possible to discriminate between different 
approaches describing the $\bar{K}N$ interaction.

The $\pi\Sigma$ mass spectrum was calculated for kaon incident momentum $p^{cm}_{K^{-}}=50-250$ MeV/c. 
However, the kinematical effects are important at low momenta and the signal of $\Lambda(1405)$ is 
masked, we have found that within our model for momenta above the 250 MeV/c, a clear bump produced 
by $\Lambda(1405)$ resonance appear in the $K^{-}+\,^{3}\mathrm{He}\rightarrow\pi\Sigma{d}$ cross 
section in the energy region between the $\bar{K}N$ and $\pi\Sigma$ thresholds, which strongly suggests 
that the clear signals of $\Lambda(1405)$ resonance should be detected by measuring of $\pi\Sigma$ 
invariant mass distributions at the relevant energies. 

By performing the fully coupled-channel calculations for $\bar{K}Nd-\pi\Sigma{d}$ system, we studied 
the dependence of the $\pi\Sigma$ mass spectrum on the $\tau_{\pi\Sigma\rightarrow\pi\Sigma}$ amplitude. 
It was shown that the full coupled-channel calculations can produce a considerably different mass 
spectrum and the inclusion of the $\tau_{\pi\Sigma\rightarrow\pi\Sigma}$ amplitude is important for 
an exact study of the $\Lambda(1405)$ resonance structure.

This work has been financially supported by the research deputy of Shahrekord 
University. The grant number was 141/2843.

\end{document}